\documentclass[12pt]{article}
\usepackage{amssymb,amsmath,epsfig,mathtools,caption}

\begin{document}

\title{\bf Accretion onto Some Well-Known Regular Black Holes}
\author{Abdul Jawad$^1$ \thanks{jawadab181@yahoo.com; abduljawad@ciitlahore.edu.pk}
and M.Umair Shahzad$^{1,2}$ \thanks{m.u.shahzad@ucp.edu.pk} \\
\small $^1$Department of Mathematics, COMSATS Institute of Information\\
\small Technology, Lahore-54000, Pakistan.\\
\small $^2$CAMS, UCP Business School, University of Central Punjab,\\
\small  Lahore, Pakistan}

\date{}
\maketitle

\begin{abstract}
In this work, we discuss the accretion onto static spherical
symmetric regular black holes for specific choices of equation of
state parameter. The underlying regular black holes are charged
regular black hole using Fermi-Dirac Distribution, logistic
distribution, nonlinear electrodynamics, respectively and
Kehagias-Sftesos asymptotically flat regular black hole. We obtain
the critical radius, critical speed and squared sound speed during
the accretion process near the regular black holes. We also study
the behavior of radial velocity, energy density and rate of change
of mass for each regular black holes.
\end{abstract}

\section{Introduction}

At present, type 1a supernova \cite{1}, Cosmic microwave background
(CMB) radiation \cite{2} and large scale structure \cite{3,4} have
shown that our universe is currently in accelerating expansion
period. Dark energy is responsible for this acceleration and it has
strange property that violates the null energy condition (NEC) and
weak energy condition (WEC) \cite{5,6} and produces strong repulsive
gravitational effects. Recent observations suggests that
approximately 74\% of our universe is occupied by dark energy and
the rest 22\% and 4\% is of dark matter and ordinary matter
respectively. Nowadays dark energy is the most challenging problem
in astrophysics. Many theories have been proposed to handle this
important problem in last two decade. Dark energy is modeled using
the relationship between energy density and pressure by perfect
fluid with equation of state (EoS) $\rho = \omega p$. The candidates
of dark energy are phantom like fluid $(\omega < -1)$, quintessence
$(-1 < \omega < -1/3)$, cosmological constant $(\omega = -1)$
\cite{7}. Other
models are also proposed for explanation of dark energy like
k-essence, DBI-essence, Hessence, dilation, tachyon, chaplygin gas
etc \cite{8}-\cite{16}.

On the other hand, existence of essential singularities (which leads
to various black holes (BHs)) is one of the major problems in
general relativity (GR) and it seems to be a common property in most
of the solutions of Einstein field equations. To avoid these
singularities, regular BHs (RBHs) have been developed. These BHs are
solutions of Einstein Equation with no essential singularity hence
their metric is regular everywhere. Strong energy condition (SEC) is
violated by these RBHs somewhere in space time \cite{17,18} while
some of these satisfy the WEC. However, it is necessary for those
RBHs to satisfy WEC having de sitter center. The study of RBHs
solutions is very important to understand the gravitational
collapse. Since Penrose cosmic censorship conjecture claims that is
singularities predicted by GR \cite{19,20} occur, they must be
explained by event horizons. Bardeen \cite{21} has done a poineer
work in this way by presenting the first RBH known as "Bardeen Black
Hole", which satisfy the WEC.

The discussion about the properties of BHs have led many interesting
phenomenon. Accretion onto the BHs is one of them. When a massive
condensed object (e.g. black holes, neutron stars, stars etc.) try
to capture a particle of the fluid from its surroundings, then the
mass of condensed object has been effected. This process is known as
accretion of fluid by condensed object. Due to accretion the planets
and star form inhomogeneous regions of dust and gas. Supermassive
BHs exist at center of giant galaxies which suggests that they could
have formed through accretion process. It is not necessary that the
mass of BH increases due to accretion process, sometimes in falling
matter is thrown away like cosmic rays \cite{22}. First of all, the
problem of accretion on compact object was investigated by Bondi
using Newtonian theory of gravity \cite{23}. After that many
researchers such as Michel \cite{24}, Babichev et al. \cite{25,26},
Jamil \cite{27} and Debnath \cite{31} have discussed the accretion
on Schwarzschild BH under different aspects. Kim et al \cite{29} and
Madrid et al. \cite{30} studied accretion of dark energy on static
BH and Kerr-Newman BH. Sharif and Abbas \cite{28} discussed the
accretion on stringy charged BHs due to phantom energy.

Recently, framework of accretion on general static spherical
symmetric BHs has been presented by Bahamonde and Jamil \cite{22}.
We have extended this general formalism for some RBHs. We analyze
the effect of mass of RBH by choosing different values of EoS
parameter. This paper is established as follows: In section
\textbf{2}, we derive general formalism for spherically static
accretion process. In section \textbf{3}, we discuss some RBHs and
for each case, we explain the critical radius, critical points,
speed of sound, radial velocities profile, energy density and rate
of change of BH mass. In the end, we conclude our results.

\section{General Formalism For Accretion}

The generalized static spherical symmetry is characterized by the
following line element
\begin{equation}\label{1}
ds^{2}=-X(r)dt^{2}+\frac{1}{Y(r)}dr^{2}
+Z(r)(d\theta^{2}+\sin\theta^{2}d\phi^{2}),
\end{equation}
where $X(r)>0$, $Y(r)>0$ and $Z(r)>0$ are the functions of $r$ only.
The energy-momentum tensor is considered as perfect fluid which is
isotropic and inhomogeneous and defined as follows
\begin{equation}\label{2}
T_{\mu\nu} = (\rho + p)u_\mu u_\nu + p g_{\mu\nu},
\end{equation}
where \emph{p} is pressure, $\rho$ is energy density and $u^{\mu}$
is the four velocity which is given by
\begin{equation}\label{3}
u^{\mu} = \frac{dx^{\mu}}{d\tau} = (u^{t},u^{r},0,0),
\end{equation}
where $\tau$ is the proper time. We have $u^{\theta}$ and $u^{\phi}$
both equal to zero due to spherical symmetry restrictions. Here
pressure, energy density and four velocity components are only the
functions of $r$. The normalization condition of four velocity must
satisfy $u^{\mu} u_{\mu} = -1$, we get
\begin{equation}\label{4}
u^{t} := \frac{dt}{d\tau} = \sqrt{\frac{u^2+Y}{XY}},
\end{equation}
where $ u = dr/d\tau = u^{r}$\cite{22}, $u^{t}$ can be negative or
positive due to square root which represents the backward or forward
in time conditions. However, $u < 0$ is required for accretion
process otherwise for any outward flows $u > 0$. Both inward and
outward flows are very important in astrophysics. One can assume
that the fluid is dark energy or any kind of dark matter. For
spherical symmetric BH, the proper dark energy model could be
obtained by generalizing Michel's theory. In dark energy accretion,
Babichev et al. \cite{25} have introduced the above generalization
on Schwarzschild black hole. Similarly, some authors \cite{22,31}
have extended this procedure for generalized static spherically
symmetric BH. In these works, equation of continuity plays an
important role which turns out to be
\begin{equation}\label{5}
(\rho + p)u\frac{X(r)}{Y(r)}\sqrt{u^2+Y(r)}Z(r) = A_{0},
\end{equation}
where $A_0$ is the constant of integration. Using $u_{\mu}
T^{\mu\nu}_{\mu} = 0$, we obtain continuity (or relativistic energy
flux) equation
\begin{equation}\label{6}
u^{\mu} \rho_{,\mu} + (\rho + p)u^{\mu}_{;\mu}=0.
\end{equation}
Furthermore, assuming $p = p(\rho)$ a certain EoS in this case.
After some calculations, the above equation becomes
\begin{equation}\label{7}
\frac{\rho^{'}}{\rho + p} + \frac{u^{'}}{u} +\frac{X^{'}}{2X} +
\frac{Y^{'}}{2Y}+ \frac{Z^{'}}{Z}=0,
\end{equation}
here prime represents the derivative with respect to $r$. By
integrating the last equation, we obtain
\begin{equation}\label{8}
uZ(r)\sqrt{\frac{X(r)}{Y(r)}}
e^{\int{\frac{d\rho}{\rho+p(\rho)}}}=-A_1,
\end{equation}
where $A_1$ is the constant of integration. By equating
Eqs.(\ref{5}) and (\ref{8}), we get
\begin{equation}\label{9}
(\rho+p)\sqrt{\frac{X(r)}{Y(r)}}\sqrt{u^2+Y}
e^{-\int{\frac{d\rho}{\rho+p(\rho)}}}=-\frac{A_0}{A_1}=A_3,
\end{equation}
where $A_3$ is another constant depends upon $A_0$ and $A_1$.
Moreover, the equation of mass flux yields
\begin{equation}\label{10}
\rho u\sqrt{\frac{X(r)}{Y(r)}}Z(r) = A_2,
\end{equation}
where $A_2$ is the constant of integration. By using Eqs.(\ref{5})
and (\ref{10}), we obtain the following important relation
\begin{equation}\label{11}
\frac{(\rho+p)}{\rho} \sqrt{\frac{X(r)}{Y(r)}}\sqrt{u^2+Y} =
\frac{A_1}{A_2}\equiv A_4,
\end{equation}
where $A_4$ is arbitrary constant which depends on $A_1$ and $A_2$.
Differentials of Eqs.(\ref{10}) and (\ref{11}) and some manipulation
leads to
\begin{equation}\label{12}
\left(V^{2} - \frac{u^2}{u^2+Y}\right)\frac{du}{u} +
\left(\left(V^{2}-1\right)\left(\frac{X^{'}}{X}-\frac{Y^{'}}{Y}\right)
+\frac{Z^{'}}{Z}V^{2}-\frac{Y^{'}}{2\left(u^{2}+Y\right)}\right)dr =
0.
\end{equation}

In addition, we have introduced the variable
\begin{equation}\label{13}
V^{2} \equiv \frac{d\ln{\rho + p}}{d\ln{\rho}} - 1.
\end{equation}
If the bracketed terms in Eq.(\ref{12}) vanishes, we obtain the
critical point (where speed of sound equal to speed of flow) which
is located at $r = r_c$. Hence at critical point, we get
\begin{equation}\label{14}
V_c^{2} = \frac{u_c^2}{u_c^2+Y(r_c)},
\end{equation}
and Eq.(\ref{12}) turns out to be
\begin{equation}\label{15}
\left(V^{2}_c-1\right)\left(\frac{X^{'}(r_c)}{X(r_c)}-\frac{Y^{'}(r_c)}{Y(r_c)}\right)+\frac{Z^{'}(r_c)}{Z(r_c)}V^{2}_{c}
= \frac{Y^{'}(r_c)}{2\left(u^{2}_c+Y(r_c)\right)}.
\end{equation}
Also, $u_c$ is the critical speed of flow evaluated at critical
value $r = r_c$. We can decoupled the above two equations and obtain
\begin{equation}\label{16}
u^{2}_{c} = \frac{Y(r_c)Z(r_c)X^{'}(r_c)}{2X(r_c)Z^{'}(r_c)},\quad
V^{2}_{c}=\frac{Z(r_c)X^{'}(r_c)}{2X(r_c)Z^{'}(r_c)+Z(r_c)X^{'}(r_c)}.
\end{equation}
The speed of sound is evaluated at $r = r_c$ as follows
\begin{equation}\label{18}
c^{2}_s=\frac{dp}{d\rho}|_{r
=r_c}=A_4\sqrt{\frac{Y(r_c)}{X(r_c)(u^{2}_c+Y(r_c))}}-1,
\end{equation}
Obviously, $u^{2}_c$ and $V^{2}_c$ can never be negative and hence
\begin{equation}\label{19}
\frac{X^{'}(r_c)}{Z^{'}(r_c)} > 0,
\end{equation}

Moreover, the rate of change of BH mass can be defined as follows
\cite{31}
\begin{equation}\label{20}
\dot{M}_{acc} = 4 \pi A_3 M^{2} (\rho + p),
\end{equation}
Here dot is derivative with respect to time. We can observe that the
mass of BH will increase for the fluid $\rho+p>0$ and hence the
accretion occurs outside the BH. Otherwise, for $\rho+p<0$ like
fluid, the mass of BH will decrease. The mass of BH cannot remain
fixed because it will decrease in hawking radiation while it will
increase in accretion. If we consider the time dependence of BH
mass, then first assume that it will not change the geometry and
symmetry of space time. Hence the space time metric remain static
spherical symmetric \cite{22}.

\section{Spherical Symmetric Metrics with Charged RBHs}

In this section, we discuss the spherically symmetric metrics with
charged RBHs in which $X(r) = Y(r)$. For this assumption, Equation
(\ref{16}) give
\begin{eqnarray}\label{21}
u^{2}_{c} = \frac{Z(r_c)X^{'}(r_c)}{2Z^{'}(r_c)},\quad
V^{2}_{c}=\frac{Z(r_c)X^{'}(r_c)}{2X(r_c)Z^{'}(r_c)+Z(r_c)X^{'}(r_c)}.
\end{eqnarray}
Although, our focus on charged RBHs metrics with event horizons, the
present analysis is forbidden for horizon space time. In many cases,
we concern on critical values (critical radius), critical
velocities, speed of sound in fluid, behavior of energy density of
fluid, radial velocity and rate of change of mass of accreting
objects. So horizon does not involve anywhere \cite{22}.

\subsection{Charged RBH Using Fermi-Dirac Distribution}

The said RBH solution has the following metric functions \cite{32}
\begin{equation}\label{1a}
X(r)= 1-\frac{2M}{r}\left(\frac{\xi(x)(\beta
r)}{\xi_{\infty}}\right)^{\beta}=Y(r),
\end{equation}
where the Fermi-Dirac distribution function is
\begin{equation}\label{1b}
\xi(x)=\frac{1}{e^{x}+1}.
\end{equation}
By replacing $x = \frac{q^2}{M\beta r}$, we can obtain the
distribution function as
\begin{equation}\label{1c}
\xi(\beta r)=\frac{1}{e^{\frac{q^2}{M\beta r}}+1},
\end{equation}
with normalization factor is $\xi_{\infty}=\frac{1}{2}$. Also the
distribution function satisfies
\begin{equation}\label{1d}
\frac{\xi(r)}{\xi_{\infty}}\rightarrow 1,
\end{equation}
where $r\rightarrow \infty$. Hence the metric functions turn out to
be
\begin{equation}\label{31a}
X(r)= Y(r)= 1-\frac{2M}{r} \left(\frac{2}{
e^{\frac{q^2}{Mr}}+1}\right)^{\beta}, Z(r)=r^2.
\end{equation}
If we set $\beta$ $\rightarrow 0$ and $\beta$ $\rightarrow$
$\infty$, we obtain
\begin{eqnarray}\label{31b}
X(r)&=& Y(r)= 1-\frac{2M}{r} e^{\frac{-q^2}{Mr}},\\\label{31c}
X(r)&=& Y(r)= 1-\frac{2M}{r} e^{\frac{-q^2}{2Mr}}.
\end{eqnarray}
In both equations, the difference of factor $2$ must be noted
\cite{32}.

It is possible to integrate the conversation laws and obtain
analytical expressions of the physical parameters. For simplicity,
we will study the barotropic case where the fluid has an equation
$p(r) = \omega \rho(r)$. Using (\ref{5}) and (\ref{11}), we obtain
\begin{equation}\label{32}
u(r)= \frac{ \left(\left(2M\left(\frac{1}{2}
e^{\frac{q^2}{Mr}}+\frac{1}{2}\right)^{-\beta}-r\right)
\left(\omega+1\right)^{2} +A^{2}_4 r \right)^{1/2}}{(\omega
+1)\sqrt{r}},
\end{equation}
\begin{figure} \centering
\epsfig{file=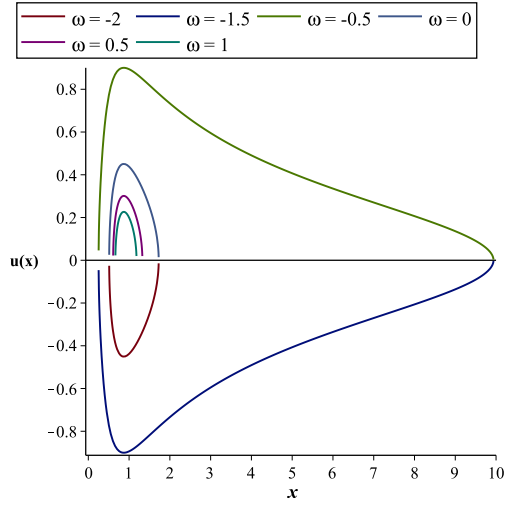,width=.5\linewidth}\caption{Velocity profile
against $x = \frac{r}{M}$ for $\beta = 1$, $q = 1.055M$, $M = 1$ and
$A_4 = 0.45$ of charged RBH using Fermi-Dirac distribution.}
\end{figure}
\begin{equation}\label{33}
\rho(r)= \frac{A_2
(\omega+1)}{r^{3/2}\sqrt{\left(2M\left(\frac{1}{2}
e^{\frac{q^2}{Mr}}+\frac{1}{2}\right)^{-\beta}-r\right)
\left(\omega+1\right)^{2} +A^{2}_4 r }}.
\end{equation}
\begin{figure} \centering
\epsfig{file=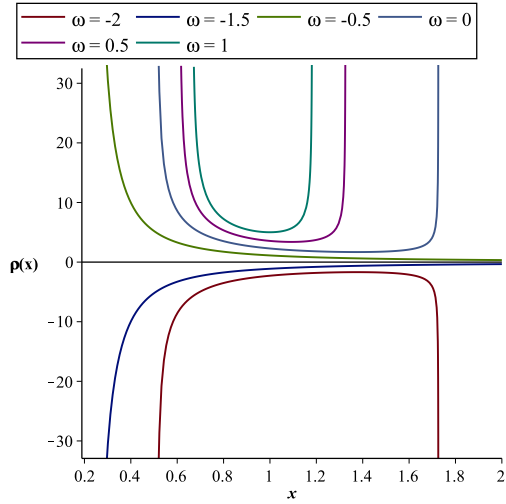,width=.5\linewidth}\caption{Energy density
against $x = \frac{r}{M}$ for $\beta = 1$, $q = 1.055M$, $M = 1$,
$A_2 = 1$ and $A_4 = 0.45$ of charged RBH using Fermi-Dirac
distribution.}
\end{figure}

The velocity profile for different values of $\omega$ is shown in
Figure \textbf{1}. Here $\omega=1,~0,~-1$ refer to stiff, dust and
cosmological constant respectively and $-1 < \omega < -1/3$ and
$\omega < -1$ refer to quintessence and phantom energy. It can be
seen that for $\omega=-1.5,~-2$ the radial velocity of the fluid is
negative and it is positive for $\omega=-0.5,~0,~0.5,~1$. If the
flow is outward then $u < 0$ is not allowed and vice versa. In the
case of $\omega=-1.5,~-0.5$ the fluid is at rest at $x = 10$. Figure
\textbf{2} represents the behavior of energy density of fluids in
the surrounding area of BH. Obviously the WEC and DEC satisfied by
dust, stiff and quintessence fluids. When phantom fluid
($\omega=-1.5,~-2$) moves towards BH then energy density decreases
and reverse will happen for  dust, stiff and quintessence fluids
($\omega=-0.5,~0,~0.5,~1$). Asymptotically $\rho~\rightarrow~0$ at
infinity for $\omega=-1.5,~-0.5$ while it approaches to maximum at
$x=1.2,~1.3,~1.8$ and near the BH.

Using this metric, Eqs.(\ref{20}) and (\ref{33}), the rate of change
of mass of RBH due to accretion becomes
\begin{equation}\label{34}
\dot{M} = \frac{4 \pi A^{2}_2 A_4
(\omega+1)}{r^{3/2}\sqrt{\left(2M\left(\frac{1}{2}
e^{\frac{q^2}{Mr}}+\frac{1}{2}\right)^{-\beta}-r\right)
\left(\omega+1\right)^{2} +A^{2}_4 r }}.
\end{equation}
\begin{figure} \centering
\epsfig{file=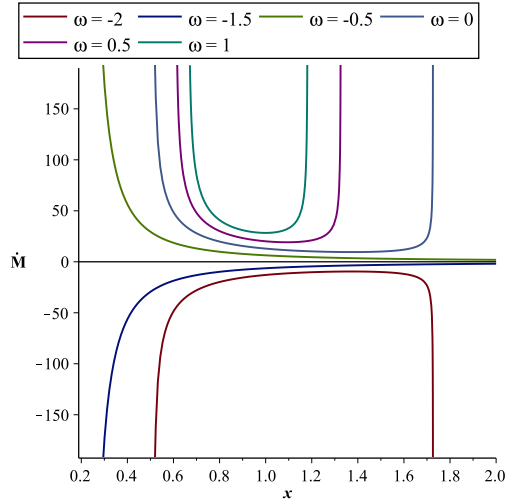,width=.5\linewidth}\caption{Rate of change of
mass of RBH against $x = \frac{r}{M}$ for $\beta = 1$, $q = 1.055M$,
$M = 1$, $A_2 = 1$ and $A_4 = 0.45$ of charged RBH using Fermi-Dirac
distribution.}
\end{figure}
Figure \textbf{3} represents the change in BH mass for different
values of $\omega$. The mass of the BH will increase near it and at
$x=1.2,~1.3,~1.7$ for $\omega=1,~0.5,~0$ respectively. On the other
hand, mass of BH decrease near it and at $x=1.7$ for $\omega=-2$.
Hence the mass of BH increases due to accretion of quintessence,
dust and stiff matter while it decreases due to accretion phantom
like fluids.

\begin{table}[h]
\begin{center}
\begin{tabular}{|c|c|c|c|}
\hline
$\omega$ & $r_c$ & $u(r_c)$ & $c^{2}_{s}$ \\
\hline
-2 & 1.37495 & 0.3138832070 & .0000002580 \\
-1.5 & 7.5044 & 0.2382908936 & -.4999997476 \\
-0.5 & 7.5044 & -0.2382908936 & -.4999997476 \\
0 & 1.3749 & -0.3138832070 & .0000002580 \\
0.5 & 1.092 & -0.2476468259 & 0.503110174\\
1 & 0.999 & -.1998921298 & 1.002986469\\
\hline
\end{tabular}
\end{center}
\caption{Charged RBH using Fermi-Dirac distribution.} \label{01}
\end{table}
The critical values, critical velocities and speed of sound are
obtained for different values of the EoS parameter in Table
\textbf{1}. Critical radius is shifting to left when $\omega\geq 0$
increases. Thus, the infalling the fluid acquires supersonic speeds
closer to BH. Same critical radius is obtained for $\omega = -2,~0$
and $\omega = -1.5, ~-0.5$ with same critical velocities but
opposite direction. We get negative speed of sound at $x = 7.5044$
and positive speed of sound for remaining critical radius. Also, the
speed of sound increases near the RBH. For this metric, we find that
\begin{equation}\label{35a}
u^{2}_c = \frac{2^{\beta - 1}\left(\left(M
r-q^{2}\right)e^{\frac{q^{2}}{\beta M
r}}+Mr\right)}{r^{2}\left(e^{\frac{q^{2}}{\beta M
r}}+1\right)^{\beta +1}},
\end{equation}
\begin{equation}\label{36}
V^{2}_c = \frac{2^{\beta}\left(\left(M
r-q^{2}\right)e^{\frac{q^{2}}{\beta M
r}}+Mr\right)}{2r^{2}\left(e^{\frac{q^{2}}{\beta M
r}}+1\right)^{\beta +1}-2^{\beta}\left(\left(3M
r+q^{2}\right)e^{\frac{q^{2}}{\beta M r}}+3Mr\right)}.
\end{equation}
Also, the condition (\ref{19}) yields
\begin{equation}\label{35a}
\frac{2^{\beta }\left(\left(M r-q^{2}\right)e^{\frac{q^{2}}{\beta M
r}}+Mr\right)}{r^{4}\left(e^{\frac{q^{2}}{\beta M
r}}+1\right)^{\beta +1}}>0.
\end{equation}

\subsection{Charged RBH Using Logistic Distribution}

The Logistic distribution function is \cite{32}
\begin{equation}\label{1aa}
\xi(x)=\frac{e^{-x}}{(e^{-x}+1)^2},
\end{equation}
in which replace $x=\frac{2q^2}{M\beta r}$ then we obtain the
distribution function
\begin{equation}\label{1bb}
\xi(\beta r)=\frac{e^{\frac{-2q^2}{M\beta
r}}}{\left(e^{\frac{-2q^2}{M\beta r}}+1\right)^2},
\end{equation}
with normalization factor is $\sigma_{\infty}=\frac{1}{4}$. Also the
distribution function satisfies
\begin{equation}\label{1cc}
\frac{\xi(r)}{\xi_{\infty}}\rightarrow 1,
\end{equation}
where $r\rightarrow \infty$. The horizons can be obtained for
$\beta=1$ where $q = 1.055M$. The metric function can be written as
\begin{equation}\label{35aa} X(r)= Y(r)= 1-\frac{2M}{r} \left(\frac{4
e^{-\sqrt{\frac{2q^{2}} {\beta M r}}}}{
\left(e^{-\sqrt{\frac{2q^2}{\beta M r}}}+1\right)^{2}}
\right)^{\beta}, Z(r)=r^2.
\end{equation}
If we set $\beta \rightarrow 0$ then we obtain the Schwarzschild BH
and if we set $\beta \rightarrow \infty$ we get
\begin{equation}\label{31aa} X(r)= Y(r)= 1-\frac{2M}{r}
e^{\frac{-q^2}{2Mr}}.
\end{equation}
It is noteworthy that this metric function corresponds to an
Ay´on-Beato and Garcýa BH \cite{32}.

The radial velocity and energy density for the metric (\ref{35aa})
using eqs. (\ref{5}) and (\ref{10}) is given by
\begin{equation}\label{36aa}
u(r)= \frac{1}{(\omega+1)\sqrt{r}}\left(\left(-2 M \left( \frac{4
e^{-\sqrt{\frac{2 q^{2}}{\beta M r}}}} {\left(e^{-\sqrt{\frac{2
q^{2}}{\beta M r}}}+1\right)^{2}}\right) ^{\beta}+r\right)\left(
\omega + 1\right)^2-A^{2}_4 r \right)^{1/2},
\end{equation}
\begin{figure} \centering
\epsfig{file=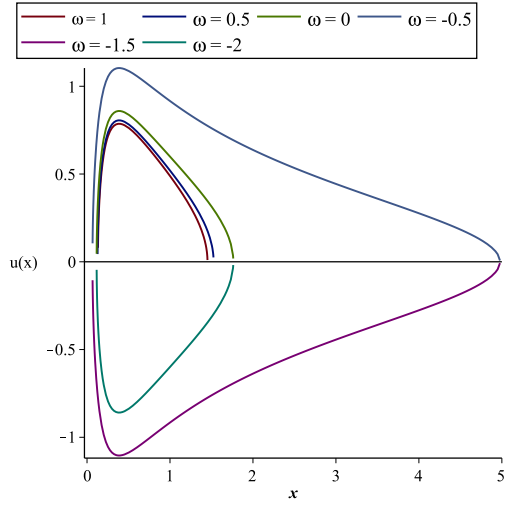,width=.5\linewidth} \caption{Velocity profile
against $x = \frac{r}{M}$ for $\beta = 1$, $q = 1.055M$, $M = 1$ and
$A_4 = 0.4$ of charged RBH using logistic distribution.}
\end{figure}
\begin{equation}\label{37aa}
\rho(r)= -\frac{A_2 (\omega+1)}{r^{3/2}\left(\left(2 M \left(
\frac{4 e^{-\sqrt{\frac{2 q^{2}}{\beta M r}}}}
{\left(e^{-\sqrt{\frac{2 q^{2}}{\beta M r}}}+1\right)^{2}}\right)
^{\beta}-r\right) \left( \omega + 1\right)^2+A^{2}_4 r
\right)^{1/2}}.
\end{equation}
\begin{figure} \centering
\epsfig{file=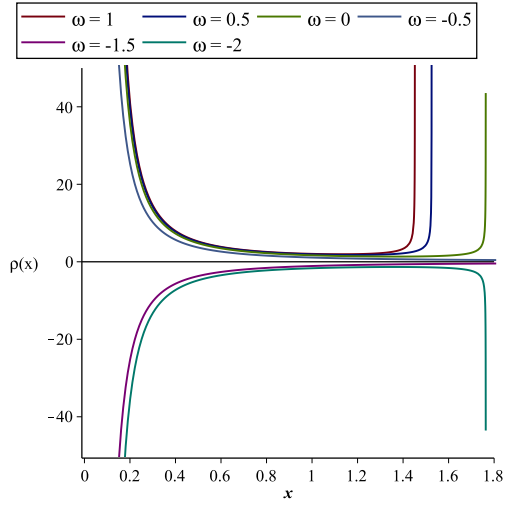,width=.5\linewidth} \caption{Energy density
against $x = \frac{r}{M}$ for $\beta = 1$, $q = 1.055M$, $M = 1$,
$A_2 = 1$ and $A_4 = 0.4$ of charged RBH using logistic
distribution.}
\end{figure}

The velocity profile for different values of $\omega$ is shown in
Figure \textbf{4}. It can be observed that for $\omega=-1.5,~-2$ the
radial velocity of the fluid is negative and it is positive for
$\omega=-0.5,~0,~1$. If the flow is inward then $u > 0$ is not
allowed and vice versa. In the case of $\omega=-2,~0$ the fluid is
at rest at $x~\approx~5$. Figure \textbf{5}, represents the behavior
of energy density of fluids in the surrounding area of BH. Obviously
the WEC and DEC are satisfied by dust, stiff and quintessence
fluids. When phantom like fluid ($\omega=-1.5,~-2$) moves towards BH
then energy density decreases and reverse will happen for dust,
stiff and quintessence fluids ($\omega=-0.5,~0,~0.5,~1$).

The $\dot{M}$ of RBH for distinct EoS parameters is obtained by
using (\ref{20})
\begin{equation}\label{38aa} \dot{M} = -\frac{4 \pi
A^{2}_2 A_4 (\omega+1)}{r^{3/2}\left(\left(2 M \left( \frac{4
e^{-\sqrt{\frac{2 q^{2}}{\beta M r}}}} {\left(e^{-\sqrt{\frac{2
q^{2}}{\beta M r}}}+1\right)^{2}}\right) ^{\beta}-r\right) \left(
\omega + 1\right)^2+A^{2}_4 r \right)^{1/2}}.
\end{equation}
\begin{figure} \centering
\epsfig{file=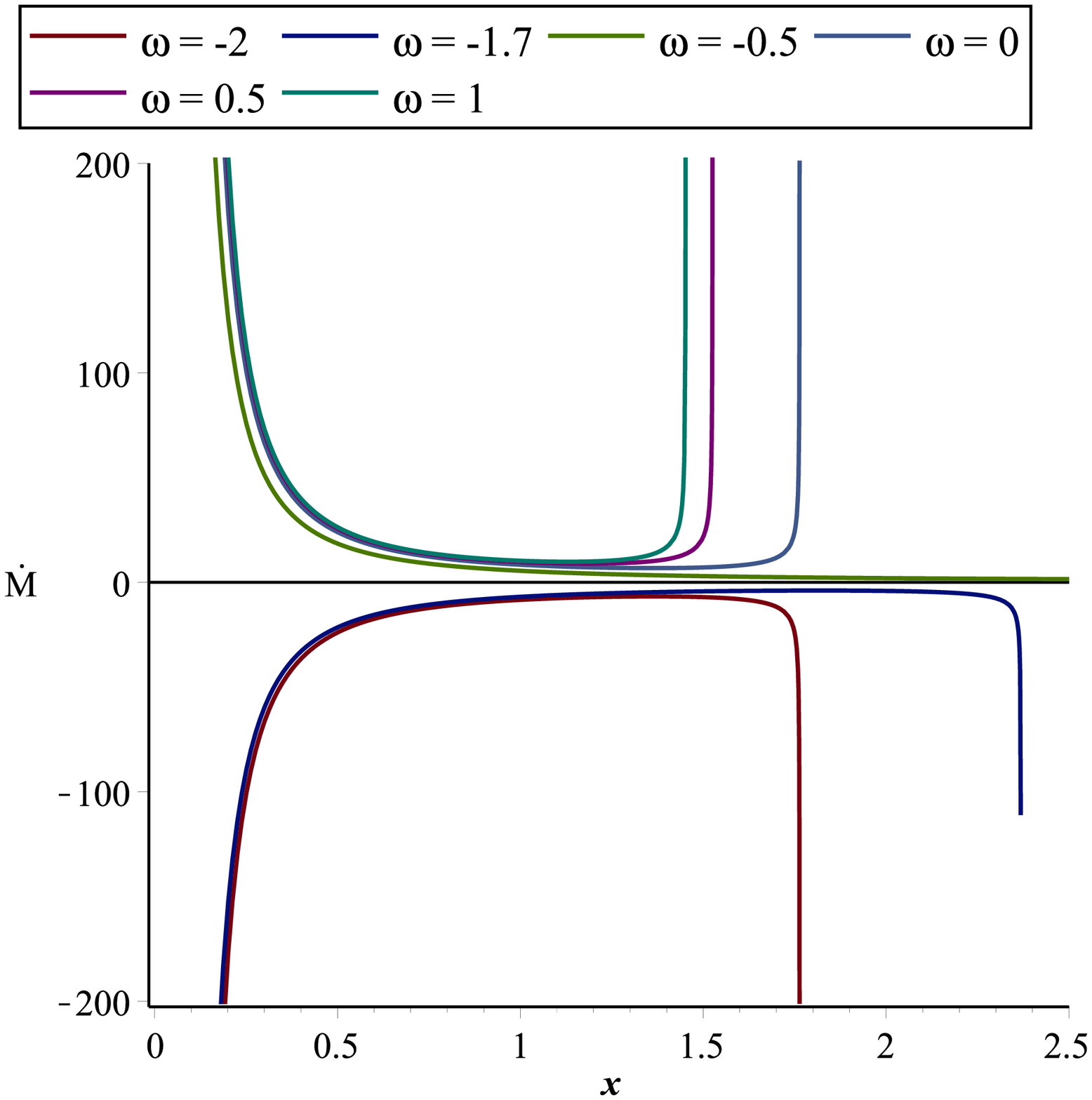,width=.5\linewidth}\caption{Rate of change of
mass of RBH against $x = \frac{r}{M}$ for $\beta = 1$, $q = 1.055M$,
$M = 1$, $A_2 = 1$ and $A_4 = 0.4$ of charged RBH using logistic
distribution.}
\end{figure}
Figure \textbf{6} represents the change in BH mass against \emph{x}.
It is evident that the mass of BH increases due to quintessence,
dust and stiff fluids and it decreases due to phantom fluids.

\begin{table}[h]
\begin{center}
\begin{tabular}{|c|c|c|c|} \hline
$\omega$ & $r_c$ & $u(r_c)$ & $c^{2}_{s}$\\
\hline
-2 & 1.36375& -0.3998729763 & -0.1018620364\\
-1.5 & 3.777412&-0.3138895411 & -0.5007414180 \\
-0.5 & 3.77412&0.3138895411 & -0.5007414180 \\
0 & 1.36375 &0.3998724197 & -0.1018620364\\
0.5 & 1.1850& 0.4018068205 & 0.116622918 \\
1 & 1.12974&0.4014558621 & 0.231766770 \\
\hline
\end{tabular}
\end{center}
\caption{Charged RBH using logistics distribution.} \label{02}
\end{table}

The critical radius, critical velocities and speed of sound are
obtained for different values of EoS parameter in Table \textbf{2}.
Critical radius is shifting to right when $\omega \geq 0$ increases.
Thus the infalling the fluid acquires supersonic speeds closer to
BH. For phantom like fluid, quintessence, dust and stiff matter the
critical radius and critical velocities are explained in the above
table. Same critical radius is obtained for $\omega=-2,~0$ and
$\omega=-1.5,~-0.5$ with same critical velocities but differ in
sign. We obtained negative speed of sound at $x=1.36375,~3.777412$
and positive speed of sound at $x=1.12974,~1.1850$. Near the BH
speed of sound will increase. For this metric we find that
\begin{equation}\label{39aa}
u^{2}_c = \frac{2^{-2+2\beta}\sqrt{M}e^{\sqrt{\frac{2q^2\beta}{Mr}}}
\left(\left(\sqrt{2}q\beta+2\sqrt{\beta M r}\right)+
\left(-\sqrt{2}q\beta+2\sqrt{\beta M
r}\right)e^{\sqrt{\frac{2q^2\beta}
{Mr}}}\right)}{\sqrt{\beta}r^{3/2}\left(e^{\sqrt{\frac{2q^2\beta}
{Mr}}}\right)},
\end{equation}
\begin{eqnarray}\label{39aa} \nonumber V^{2}_c &=& \left(2^{\beta}M
e^{\sqrt{\frac{2q^2\beta}{Mr}}}\left(\left(\sqrt{2}q\beta+2\sqrt{\beta
M r}\right)+ \left(-\sqrt{2}q\beta+2\sqrt{\beta M
r}\right)e^{\sqrt{\frac{2q^2\beta}{Mr}}}\right)\right)\\
\nonumber &&\left(4\sqrt{\beta M}r^{3/2}
\left(e^{\sqrt{\frac{2q^2\beta}{M
r}}}\right)-2^{\beta}e^{-\sqrt{\frac{2\beta q^2}{M
r}}}\left(\left(-\sqrt{2}M q \beta
+6 \sqrt{\beta r}M^{3/2}\right)\right.\right.\\
&&\left.\left.e^{-\sqrt{\frac{2q^2}{\beta M r}}}+ \sqrt{2}M q
\beta+6\sqrt{\beta r}M^{3/2}\right)\right)^{-1}.
\end{eqnarray}
Also, the condition (\ref{19}) yields
\begin{equation}\label{40aa}
\frac{2^{-1+2\beta}\sqrt{M}e^{-\sqrt{\frac{2q^2\beta}{Mr}}}
\left(\left(\sqrt{2}q\beta+2\sqrt{\beta M
r}\right)e^{-\sqrt{\frac{2q^2\beta}
{Mr}}}+\left(-\sqrt{2}q\beta+2\sqrt{\beta M r}\right)\right)}
{\sqrt{\beta}r^{7/2}\left(e^{\sqrt{\frac{2q^2\beta}{Mr}}}\right)}>0
\end{equation}

\subsection{Charged RBH from Nonlinear Electrodynamics}

Using the line element
\begin{equation}\label{31aaa}
X(r)= Y(r)= 1-\frac{2M(r)}{r}.
\end{equation}
Here the function
\begin{equation}\label{1aaa}
M(r) = M\left(1-tanh\left(\frac{q^2}{2Mr}\right)\right),
\end{equation}
and its associated electric field source is
\begin{equation}\label{1bbb}
E=\frac{q}{r^2}\left(1-tanh^{2}\left(\frac{q^2}{2Mr}\right)\right)
\left(1-\frac{q^2}{4Mr}tanh\left(\frac{q^2}{2Mr}\right)\right),
\end{equation}
where \emph{q} and \emph{M} represent electric charge and mass
respectively \cite{33}. The solution elaborate RBH and its global
structure is like R-N BH. The asymptotic behavior of the solution is
\begin{equation}\label{1}
X(r) = 1-\frac{2M}{r}+\frac{q^2}{r^2}+O\left(\frac{1}{r^4}\right).
\end{equation}
So the metric function
\begin{equation}\label{35aaa} X(r)= Y(r)=
1-\frac{2M}{r} \left(M\left(1-tanh(\frac{q^2}{2Mr})\right)\right),
Z(r)=r^2.
\end{equation}

The radial velocity and energy density for this metric are given by
\begin{equation}\label{41aaa} u(r) = \frac{\sqrt{\left(-2 M
tanh\left(\frac{q^2}{2Mr}\right)+2M-r\right)\left(\omega+1\right)^2+A^{2}_4
r}}{(\omega +1)\sqrt{r}},
\end{equation}
\begin{equation}\label{42aaa} \rho(r) = \frac{(\omega
+1)A_2}{r^{3/2}\sqrt{\left(-2 M
tanh(\frac{q^2}{2Mr}+2M-r)\right)\left(\omega+1\right)^2+A^{2}_4
r}}.
\end{equation}
\begin{figure} \centering
\epsfig{file=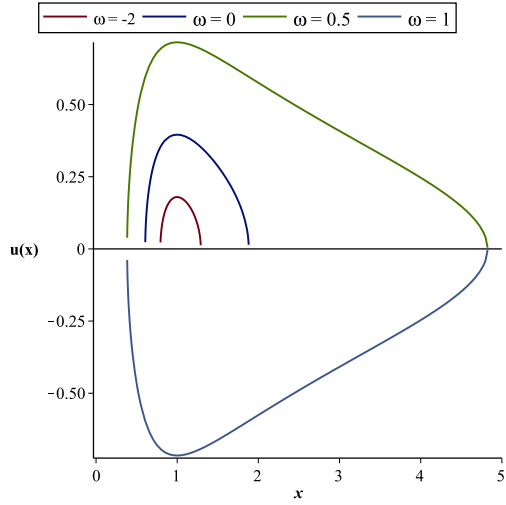,width=.5\linewidth} \caption{Velocity profile
against $x = \frac{r}{M}$ for $q = 1.055M$, $M = 1$ and $A_4 = 0.7$
of charged RBH from nonlinear electrodynamics.}
\end{figure}
\begin{figure} \centering
\epsfig{file=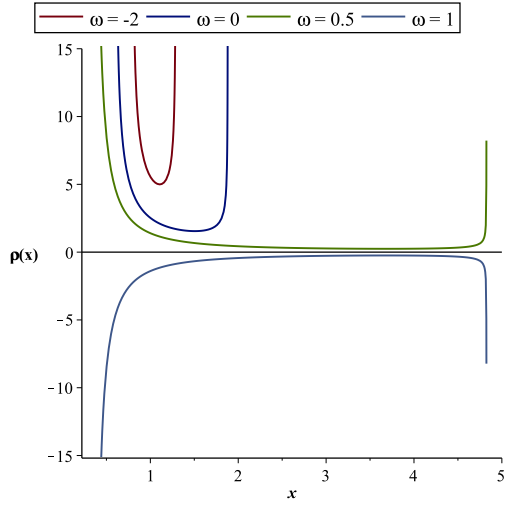,width=.5\linewidth} \caption{Energy density
against $x = \frac{r}{M}$ for $q = 1.055M$, $M = 1$, $A_2 = 1$ and
$A_4 = 0.7$ of charged RBH from nonlinear electrodynamics.}
\end{figure}

The absolute value of the velocity profile for different values of
$\omega$ is shown in Figure \textbf{7}. It can be observed that for
$\omega =  -2$ the radial velocity of the fluid is negative and it
is positive for $\omega=0.5,~0,~1$. If the flow is inward then $u >
0$ is not allowed and vice versa. In the case of $\omega = -2,~0$
the fluid is at rest at $x \approx 5$. Figure \textbf{8} represents
the energy density of fluids in the region of BH. It is apparent
that the WEC and DEC is satisfied by phantom fluids. When phantom
fluids moves towards the BH the energy density increases on the
other hand it decreases for dust and stiff matter.

The rate of change of mass is given by
\begin{equation}\label{43aaa}
\dot{M} = -\frac{4 \pi A^{2}_2 A_4 (\omega+1)}{r^{3/2}\sqrt{\left(-2
M tanh(\frac{q^2}{2Mr}+2M-r)\right)\left(\omega+1\right)^2+A^{2}_4
r}}.
\end{equation}
\begin{figure} \centering
\epsfig{file=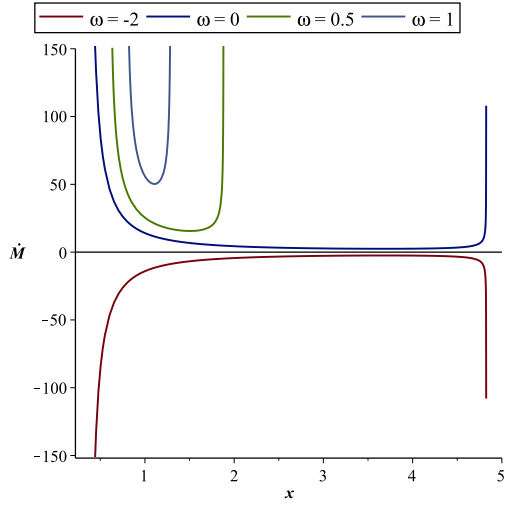,width=.5\linewidth}\caption{Rate of change of
mass of RBH against $x = \frac{r}{M}$ for $q = 1.055M$, $M = 1$,
$A_2 = 1$ and $A_4 = 0.7$ of charged RBH from nonlinear
electrodynamics.}
\end{figure}
The rate of change of in the BH mass against \emph{x} is plotted in
Figure \textbf{9}. Due to accretion of dust and stiff matter the
mass of the BH will increase for small values of \emph{x} and vice
versa for phantom fluids. It is also noted that the maximum rate of
RBH mass increases due to $\omega~=~1$ followed by
$\omega=0.5,~0,~-2$.

\begin{table}[h]
\begin{center}
\begin{tabular}{|c|c|c|c|}
\hline
$\omega$&$r_c$ & $u(r_c)$ & $c^{2}_{s}$\\
\hline
-2 & 3.685523529&-0.2993097288 &-0.125 \\
0 & 3.685523529&0.2993097288&-0.125 \\
0.5 & 1.506050868& 0.2844719573 & 0.312500584\\
1 & 1.106971797& 0.1633564212& 0.750003072 \\
\hline
\end{tabular}
\end{center}
\caption{Charged RBH from nonlinear electrodynamics.}\label{03}
\end{table}

The critical values, critical velocities and speed of sound are
obtained for different values of EoS parameter in Table \textbf{3}.
Critical radius is shifting to right when $\omega \geq 0$ increases.
Speed of sound is negative at $x = 3.685523529$ and near the BH the
speed of sound will increase. For this RBH we find that
\begin{equation}\label{39aaa}
u^{2}_c =
\frac{q^2\left(tanh^{2}\left(\frac{q^2}{2Mr}\right)-1\right)+2 M
r\left(-tanh\left(\frac{q^2}{2Mr}\right)+1\right)}{4 r^2},
\end{equation}
\begin{equation}\label{39aaa}
V^{2}_c =
\frac{q^2\left(tanh^{2}\left(\frac{q^2}{2Mr}\right)-1\right)+2 M
r\left(-tanh\left(\frac{q^2}{2Mr}\right)+1\right)}{4
r^2+q^2\left(tanh^{2}\left(\frac{q^2}{2Mr}\right)-1\right)+6+ M
r\left(tanh\left(\frac{q^2}{2Mr}\right)-1\right)}.
\end{equation}
Also, the condition (\ref{19}) yields
\begin{equation}\label{12aaa}
\frac{\left(tanh^{2}\left(\frac{q^2}{2Mr}\right)-1\right) q^{2}+2
Mr\left(-tanh\left(\frac{q^2}{2Mr}\right)+1\right)}{2 r^{4}}> 0.
\end{equation}

\subsection{Kehagias - Sftesos asymptotically flat BH}

KS studied the following BH metric
\begin{equation}\label{35dddd}
X(r)= Y(r)= 1+b r^{2}-\sqrt{b^{2}r^{4}+ 4Mbr}, Z(r)=r^2.
\end{equation}
In the frame work of Horava theory, where \emph{m} is the mass,
\emph{b} is the positive constant related to coupling constant of
theory. The metric asymptotically behaves the usual Schwarzschild BH
\cite{34}
\begin{equation}\label{1dddd}
X(r)=Y(r)\approx 1-\frac{2M}{r}+O\left(\frac{1}{r^4}\right),
\end{equation}
for $r \gg (\frac{r}{b})^{1/3}$. The KS metric have two horizons at
\begin{equation}\label{1eeee}
r_{\pm} = M\left(1\pm \sqrt{\left(1-\frac{1}{2 b
M^2}\right)}\right),
\end{equation}
with $2bM^2\geq1$ \cite{34}.

The radial velocity and energy density are given by
\begin{equation}\label{40jjjj}
u(r) = \frac{\left(A^{2}_4+\left(\sqrt{b^{2}r^{4}+
4Mbr}-br^{2}-1\right)\left(\omega +1\right)^{2}
\right)^{1/2}}{\omega +1},
\end{equation}
\begin{figure} \centering
\epsfig{file=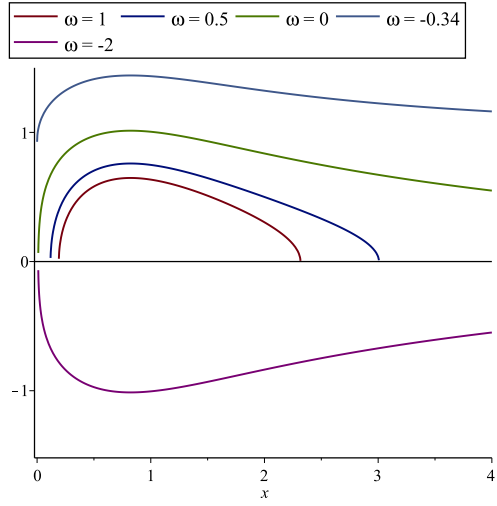,width=.5\linewidth} \caption{Velocity profile
against $x = \frac{r}{M}$ for $M = 1$, $b = 0.9$ and $A_4 = 0.9$ of
Kehagias - Sftesos asymptotically flat BH.}
\end{figure}
\begin{equation}\label{40kkkk}
\rho(r) = \frac{A_2(\omega
+1)}{r^{2}\left(A^{2}_4+\left(\sqrt{b^{2}r^{4}+4Mbr}-br^{2}-1\right)\left(\omega
+1\right)^{2}\right)^{1/2}}.
\end{equation}
\begin{figure} \centering
\epsfig{file=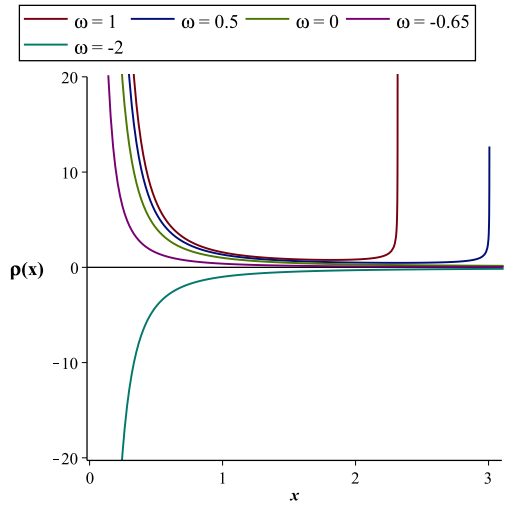,width=.5\linewidth} \caption{Energy density
against $x = \frac{r}{M}$ for $M = 1$, $A_2 = 1$, $b = 0.9$ and $A_4
= 0.9$ of Kehagias - Sftesos asymptotically flat BH.}
\end{figure}

The radial velocity for different values of $\omega$ is shown in
Figure \textbf{10}. The radial velocity is negative for phantom like
fluid and positive for quintessence, dust and stiff matter. The
evolution of energy density of fluids in the surrounding area of RBH
is plotted in Figure \textbf{11}. The energy density for phantom
fluids is negative while the energy density for stiff, dust and
quintessence fluids is positive.

For this RBH, rate of change of mass becomes
\begin{equation}\label{40aaaaaaa}
\dot{M} = \frac{4 \pi A^{2}_2 A_4(\omega
+1)}{r^{2}\left(A^{2}_4+\left(\sqrt{b^{2}r^{4}+4Mbr}-br^{2}-1\right)\left(\omega
+1\right)^{2}\right)^{1/2}}.
\end{equation}
\begin{figure} \centering
\epsfig{file=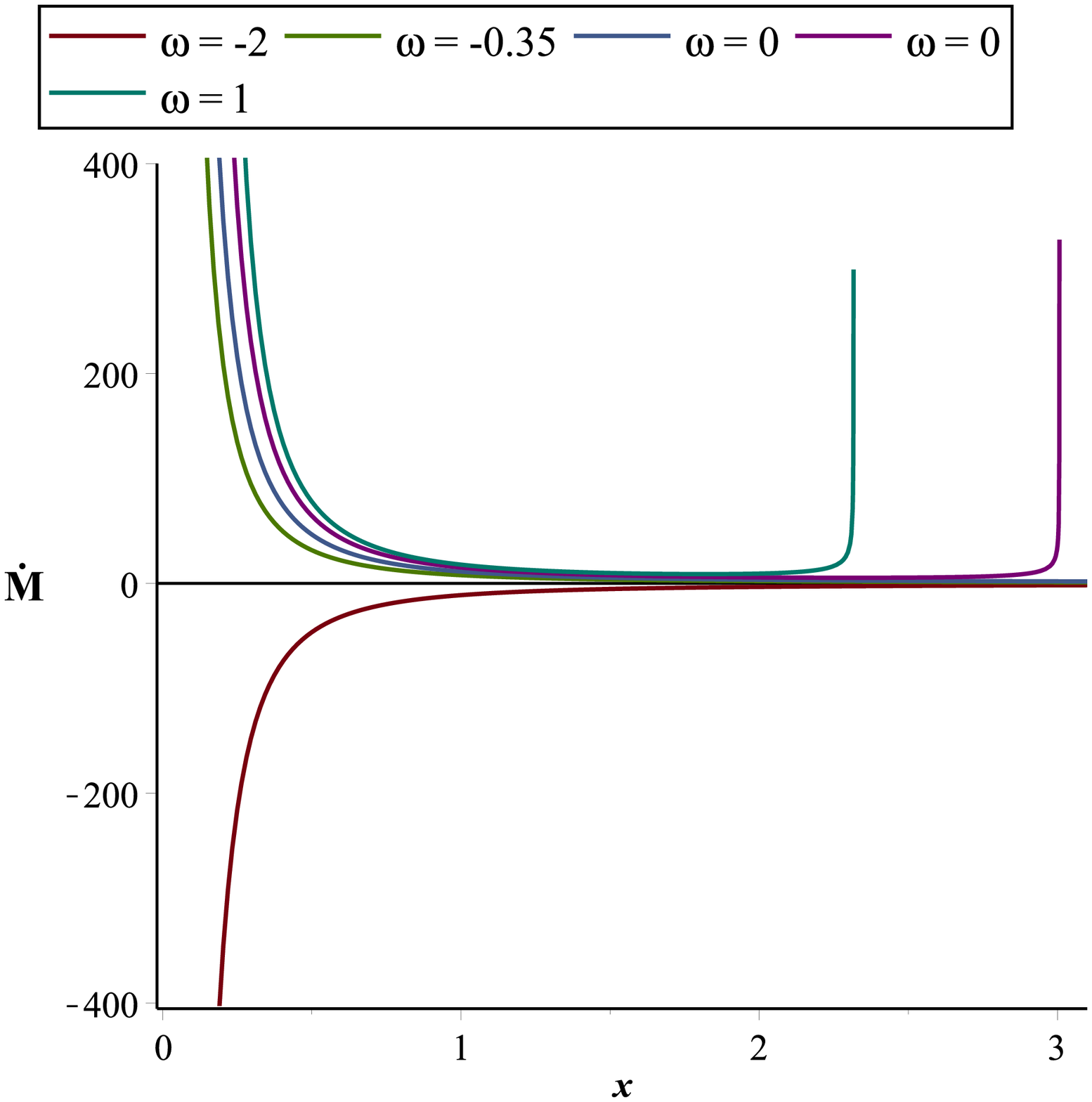,width=.5\linewidth}  \caption{Rate of change of
mass of RBH against $x = \frac{r}{M}$ for $M = 1$, $A_2 = 1$, $b =
0.9$ and $A_4 = 0.9$ of Kehagias - Sftesos asymptotically flat BH.}
\end{figure}
Figure \textbf{11} represents the rate of change in RBH mass against
\emph{x}. We see that the RBH mass will increase for $\omega=-0.35,~
0,~0.5,~1$ and it will decrease for $\omega=-2$.

\begin{table}[h]
\begin{center}
\begin{tabular}{|c|c|c|c|}
\hline
$\omega$&$r_c$& $u(r_c)$ & $c^{2}_{s}$\\
\hline
-2 & 7.8946 & -0.2505321935&0.0000008330\\
-0.34 & 30267.74 & 0.6327458490&-0.0513167023 \\
0 & 7.8946 & 0.2505321736&0.0000008330 \\
0.5 & 2.3185 & 0.3961993774&0.500013404 \\
1 & 1.8183 & 0.3888079314&0.500013404 \\
\hline
\end{tabular}
\end{center}
\caption{Kehagias - Sftesos asymptotically flat BH.}\label{04}
\end{table}

The critical values, critical velocities and speed of sound for
different values of $\omega$ is presented in the Table \textbf{4}.
For quintessence matter, we obtain very large critical radius.
Similarly as before, we obtain the same critical radius for dust and
phantom like fluids and same critical velocities but differ in sign.
If we increase the EoS parameter then the critical radius is shifted
near RBH. It is evident that the critical velocity is negative for
phantom like fluid and positive for quintessence, dust and stiff
matter. The speed of sound is negative at $x = 30267.74$ and
positive for remaining critical radius. For this metric, we find
that
\begin{equation}\label{39gggg}
u^{2}_c = \frac{r}{4}\left(2br - \frac{2b^{2} r^{3}+2 M
b}{\sqrt{r\left(b^{2} r^{3}+4 M b\right)}}\right),
\end{equation}
\begin{equation}\label{39hhhh}
V^{2}_c = \frac{r^2 \left(2br - \frac{2b^{2} r^{3}+2 M
b}{\sqrt{r\left(b^{2} r^{3}+4 M b\right)}}\right)}{r^{2}\left(2br -
\frac{2b^{2} r^{3}+2 M b}{\sqrt{r\left(b^{2} r^{3}+4 M
b\right)}}\right)+4r\left(1 + br^{2}-{\sqrt{r\left(b^{2} r^{3}+4 M
b\right)}}\right)}.
\end{equation}
The condition (\ref{19}) becomes
\begin{equation}\label{40hhhh}
\frac{\left(2br - \frac{2b^{2} r^{3}+2 M b}{\sqrt{r\left(b^{2}
r^{3}+4 M b\right)}}\right) }{2r}>0.
\end{equation}

\section{Concluding Remarks}

In this work, we have investigated the accretion onto various RBHs
(such as RBH using Fermi-Dirac Distribution, RBH using logistic
distribution, RBH using nonlinear electrodynamics and
Kehagias-Sftesos asymptotically flat RBH) which are asymptotically
leads to Schwarzschild and Reissner-Nordstrom  BHs (most of them
satisfy the WEC). We have followed the procedure of Bahamonde and
Jamil \cite{22} and obtained the critical points, critical
velocities and the behavior of sound speed for chosen RBHs.
Moreover, we have analyzed the behavior of radial velocity, energy
density and rate of change of mass for RBHs for various EoS
parameters. For calculating these quantities, we have assumed the
barotropic EoS and found the relationship between the conservation
law and barotropic EoS. We have found that the radial velocity ($u$)
of the fluid is positive for stiff, dust and quintessence matter and
it is negative for phantom-like fluids. If the flow is inward then
$u < 0$ is not allowed and $u > 0$ is not allowed for outward flow.
Also, we have obtained that the energy density remains positive for
quintessence, dust and stiff matter while becomes negative for
phantom-like fluid near RBHs.

In addition, the rate of mass of BH is dynamical quantity, so the
analysis of the nature its mass in the presence of various dark
energy models may become very interesting in the present scenario.
Also, the sensitivity (increasing or decreasing) of BHs mass depend
upon the nature of fluids which accretes onto it. Therefore, we have
considered the various possibilities of accreting fluids such as
dust and stiff matter, quintessence and phantom. We have found that
the rate of change of mass of all RBHs increases for dust and stiff
matter, quintessence-like fluid since these fluids do not have
enough repulsive force. However, the mass of all RBHs decreases in
the presence of phantom-like fluid (and the corresponding energy
density and radial velocity becomes negative) because it has strong
negative pressure. This result shows the consistency with several
works \cite{22, 31, 35,27r,28r,29r,30r}. Also, this result favor the
phenomenon that universe undergoes the big rip singularity, where
all the gravitationally bounded objects dispersed due to phantom
dark energy.

Although, we have assumed the static fluid which may be extended for
non-static fluid without assuming any EoS and can be obtained more
interesting results. This is left for future considerations.

\end{document}